\documentclass{article}

\usepackage{arxiv}

\usepackage[utf8]{inputenc} % allow utf-8 input
\usepackage[T1]{fontenc}    % use 8-bit T1 fonts
\usepackage[hidelinks]{hyperref}       % hyperlinks
\usepackage{amsmath}
\usepackage{multirow}
\usepackage{url}            % simple URL typesetting
\usepackage{booktabs}       % professional-quality tables
\usepackage{amsfonts}       % blackboard math symbols
\usepackage{nicefrac}       % compact symbols for 1/2, etc.
\usepackage{microtype}      % microtypography
\usepackage{lipsum}		% Can be removed after putting your text content
\usepackage{graphicx}
\usepackage{xspace}

\newcommand{\ie}{i.e.\xspace}
\newcommand{\eg}{e.g.\xspace}
\newcommand{\et}{et al.\xspace}
\newcommand{\rn}{{\sf BFRN}\xspace}
\newcommand{\crn}{{\sffamily cBFRN}\xspace}
\newcommand{\prn}{{\sffamily pBFRN}\xspace}
\newcommand{\irn}{{\sffamily iBFRN}\xspace}

\title{Understanding the Structure and Resilience of the Brazilian Federal Road Network Through Network Science}

\author{ \href{https://orcid.org/0000-0002-1355-7380}{\includegraphics[scale=0.06]{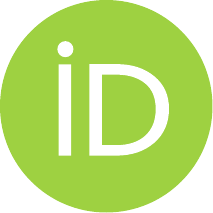}\hspace{1mm}J\'ulio Taveira}\\
	University of Pernambuco, \\
	Recife, Brazil\\
	Federal Highway Police University \\
        Florian\'opolis, Brazil\\
	\texttt{jcft@ecomp.poli.br} \\
	\And
	\href{https://orcid.org/0000-0003-1200-225X}{\includegraphics[scale=0.06]{orcid.pdf}\hspace{1mm}Fernando Buarque}\\
	University of Pernambuco, \\
	Recife, Brazil\\
	\texttt{fbln@ecomp.poli.br} \\
        \And
	\href{https://orcid.org/0000-0002-6479-6429}{\includegraphics[scale=0.06]{orcid.pdf}\hspace{1mm}Ronaldo Menezes}\\
    Department of Computer Science\\
	University of Exeter\\
	Exeter, UK\\
	\texttt{r.menezes@exeter.ac.uk} \\
}

\renewcommand{\headeright}{J. Taveira et al.}
\renewcommand{\shorttitle}{Structure and Resilience of the Brazilian Federal Road Network}

\hypersetup{
pdftitle={A template for the arxiv style},
pdfsubject={q-bio.NC, q-bio.QM},
pdfauthor={David S.~Hippocampus, Elias D.~Striatum},
pdfkeywords={First keyword, Second keyword, More},
}

\begin{document}
\maketitle

\begin{abstract}
Understanding how transportation networks work is important for improving connectivity, efficiency, and safety. In Brazil, where road transport is a significant portion of freight and passenger movement, network science can provide valuable insights into the structural properties of the infrastructure, thus helping decision makers responsible for proposing improvements to the system. This paper models the federal road network as weighted networks, with the intent to unveil its topological characteristics and identify key locations (cities) that play important roles for the country through 75,000 kilometres of roads. We start with a simple network to examine basic connectivity and topology, where weights are the distance of the road segment. We then incorporate other weights representing number of incidents, population, and number of cities in-between each segment. We then focus on community detection as a way to identify clusters of cities that form cohesive groups within a network. Our findings aim to bring clarity to the overall structure of federal roads in Brazil, thus providing actionable insights for improving infrastructure planning and prioritising resources to enhance network resilience.
\end{abstract}

% keywords can be removed

\section{Introduction}
\setcounter{footnote}{0}

In the global economy, road infrastructures are a fundamental component because they are key in facilitating the movement of goods and people. Many countries rely heavily on road networks for the transport of commodities, with some nations being more dependent than others. For instance, in geographically vast countries with dispersed urban centres, road transport becomes essential for economic activities and regional connectivity.\footnote{Five of the six largest countries in area also rank among the top five in road network size. Source: \url{https://en.wikipedia.org/wiki/List_of_countries_by_road_network_size} (accessed: 25 October 2024).}

Brazil's dependence on roads is particularly significant, serving as the primary mode of transportation for freight (and passengers) across its extensive, and often difficult access, territory. The federal highways (henceforth called ``roads'', given that they are not always highways, \ie, multiple lanes) play a crucial role in the development and integration of smaller cities. It has been shown that the efficiency and reliability of roads directly impact economic growth, social cohesion, and access to services \cite{li2017road, ferrari2018economic}.

Brazil's has a specific police force responsible for federal roads: the Federal Highway Police ({\rm Pol\'icia Rodovi\'aria Federal, PRF}). They are responsible for about 75,000 kilometres of roads out of the nearly 2 million km total~\cite{CNT2022}. While this sounds small compared to the total, they are the main arteries of road transportation in Brazil; 60\% of goods are transported using the federal road structure making its connectivity \cite{bottasso2021roads} crucial to the country.

According to the Brazilian constitution \cite{BrazilConstitution1988}, the PRF is responsible for patrolling federal roads, enforcing traffic laws, and ensuring the road safety. Their duties include accident prevention, combating criminal activities, and providing assistance to motorists, which are vital for maintaining the operational integrity of the nation's road network, among others. While the PRF is very knowledgeable about the road infrastructure, this knowledge is often distributed, not providing a holistic view of the infrastructure. By analysing network structures, authorities can identify critical nodes and links that require investment, improvement, or even more policing.

While the construction of new federal roads is relatively infrequent in Brazil (the country continues to depend heavily on infrastructure investments made in the 1960s and 1970s \cite{bottasso2021roads}, see Section \ref{sec:data}), understanding the structure of the existing network is paramount. The network's structural properties are influenced by factors such as human mobility patterns \cite{barbosa2018human}, population in neighbouring cities, and various events including festivals, religious activities, holidays, and freight logistic decisions. These factors can lead to changes in the significance and utilisation of specific roads, highlighting the importance of a comprehensive structural analysis. In this study, we model the Brazilian federal road network using the 546 largest cities as nodes; medium and large cities as per the Brazilian Geography and Statistics Institute (IBGE)\cite{ibge_censo_2022}.

\section{Transportation Networks}
\label{sec:transport}

Transportation networks are a critical aspect of modern society, and intrinsically linked to economic growth and social development. They facilitate the movement of goods, services, and people, thereby connecting markets and fostering globalisation \cite{rodrigue2020geography}. The structure and dynamics of these networks is an essential part of the process of improving efficiency, resilience, and sustainability around the world.

Over the past few decades, the world has effectively become smaller due to advancements in transportation and communication technologies. This phenomenon, often referred to as the ``time-space convergence,'' implies that the relative distance between places decreases as connectivity improves \cite{harvey2020condition}. For example, the average transatlantic travel time for freight shipments in the 1800s was about 30 days, while today is about 8 hours. 
% The same has happened within nations, a notable case is China which invested heavily in high-speed rail, reducing for instance the average distance between Beijing and Shanghai from 10 hours to 4.5 hours.
While increased connectivity offers benefits, it also presents concerns. Faster transportation networks can facilitate the rapid spread of diseases, as evidenced by the global transmission of pandemics like COVID-19 \cite{chinazzi2020effect}. Similarly, invasive species (\eg, plants, animals) can spread more easily through connected pathways, disrupting ecosystems and sometimes even local economies \cite{hulme2009trade}. 

Network Science has been instrumental in modelling transportation networks, offering tools and methods to analyse their complex structures and dynamics \cite{newman2003structure}. Numerous studies have employed network theory to investigate various modes of transportation and spatial networks~\cite{barthelemy2011spatial}, providing insights into their topology, robustness, and vulnerability. For instance, air transportation networks have been extensively studied due to their global importance \cite{guimera2005worldwide,li2016world}. Rail networks have also been a subject of interest with indications that these networks have small-world properties, identifying its structural characteristics and the implications for efficiency and robustness \cite{sen2003small,liu2007small}. In maritime transport, Kaluza \et~\cite{kaluza2010complex} studied the global cargo ship network, highlighting patterns in maritime traffic and their environmental impact.

In addition to air, rail and maritime studies, there has been significant work in urban road networks~\cite{akbarzadeh2018look,badhrudeen2022geometric,chalkiadakis2022urban} but comparatively fewer investigations into country-wide road networks \cite{weber2012evolving,tak2018measuring}, especially in the context of the Global South and large-scale countries. Road networks are particularly crucial in such regions due to their role in regional connectivity and economic development, where other forms of transport may be less developed or accessible. 

In this paper, we focus on the Brazilian Federal Road Network (\rn) using Network Science methodologies. By describing its structural properties, we aim to provide insights that can inform data-driven decisions for infrastructure development, policymaking, and strategic planning. 
% We consider several variables as weights in these networks: distance, road incidents, number of cities, and population.

\section{Data and Methods}
\label{sec:data}

Brazil is the largest country in South America and the fifth largest in the world, covering an area of approximately 8.5 million square kilometres. Due to its vast size and diverse geography, efficient transportation infrastructure is essential for economic development and national integration. However, Brazil lacks connectivity via rail and often many locations are not connected via airports, which make the road network crucial to many of the country's regions.

The federal roads in Brazil are under the jurisdiction of the federal government and serve as the main arteries linking major cities, ports, airports, strategic areas, and even to neighbouring countries. The primary purpose of these roads is to promote national connectivity, support economic activities, and ensure access to all areas of the country.

The development of the Brazilian road network happened mostly in mid-20th century, particularly during the 1950s and 1960s~\cite{skidmore2009brazil}. Brazil investment can be supported by economic theories which suggest that improved transportation infrastructure leads to reduced product costs, increased trade, and regional development \cite{aschauer1989public}.

\begin{figure}[ht]
    \centering
    \includegraphics[width=0.5\linewidth]{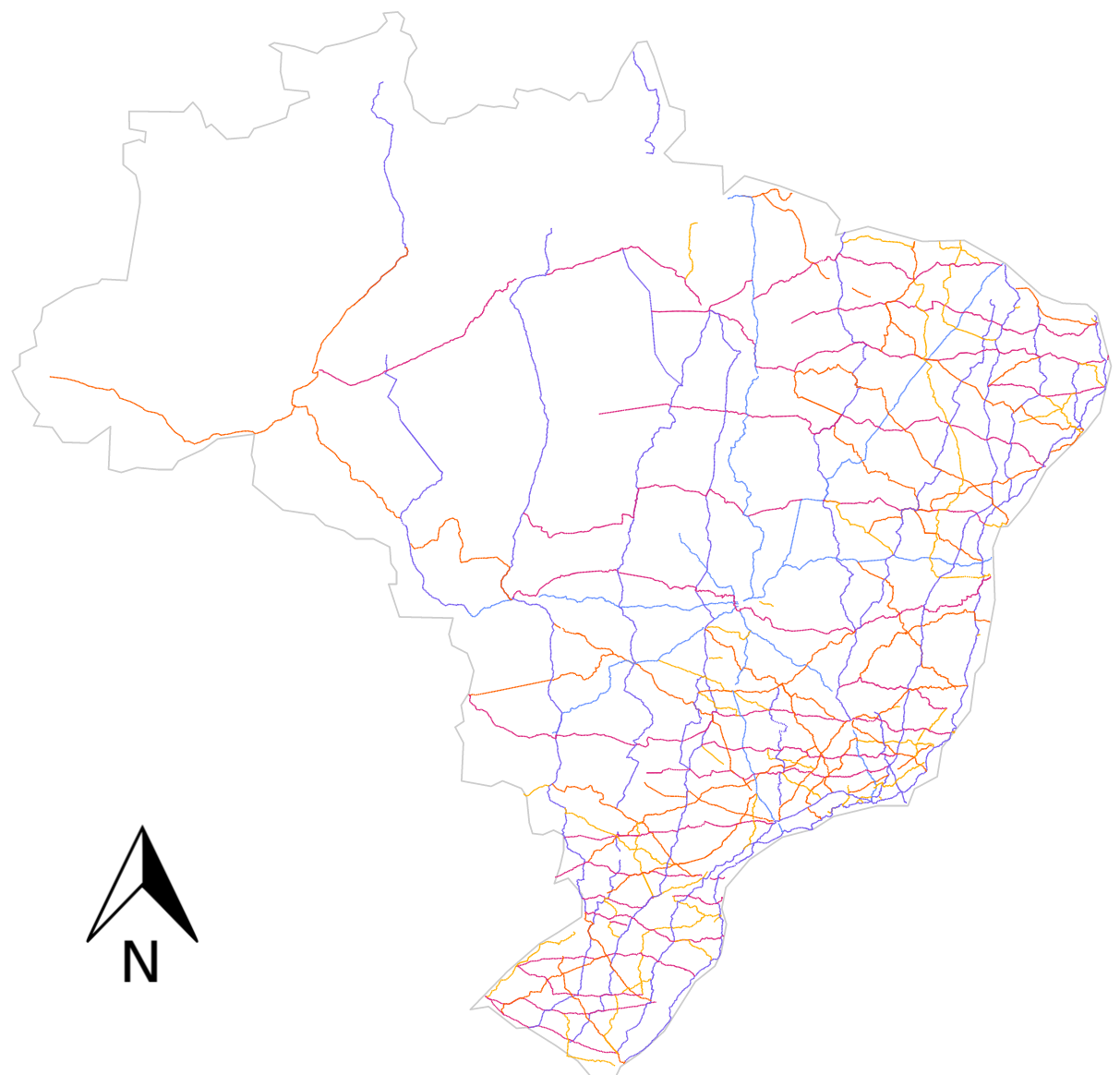}\hfill
    \includegraphics[width=0.48\linewidth]{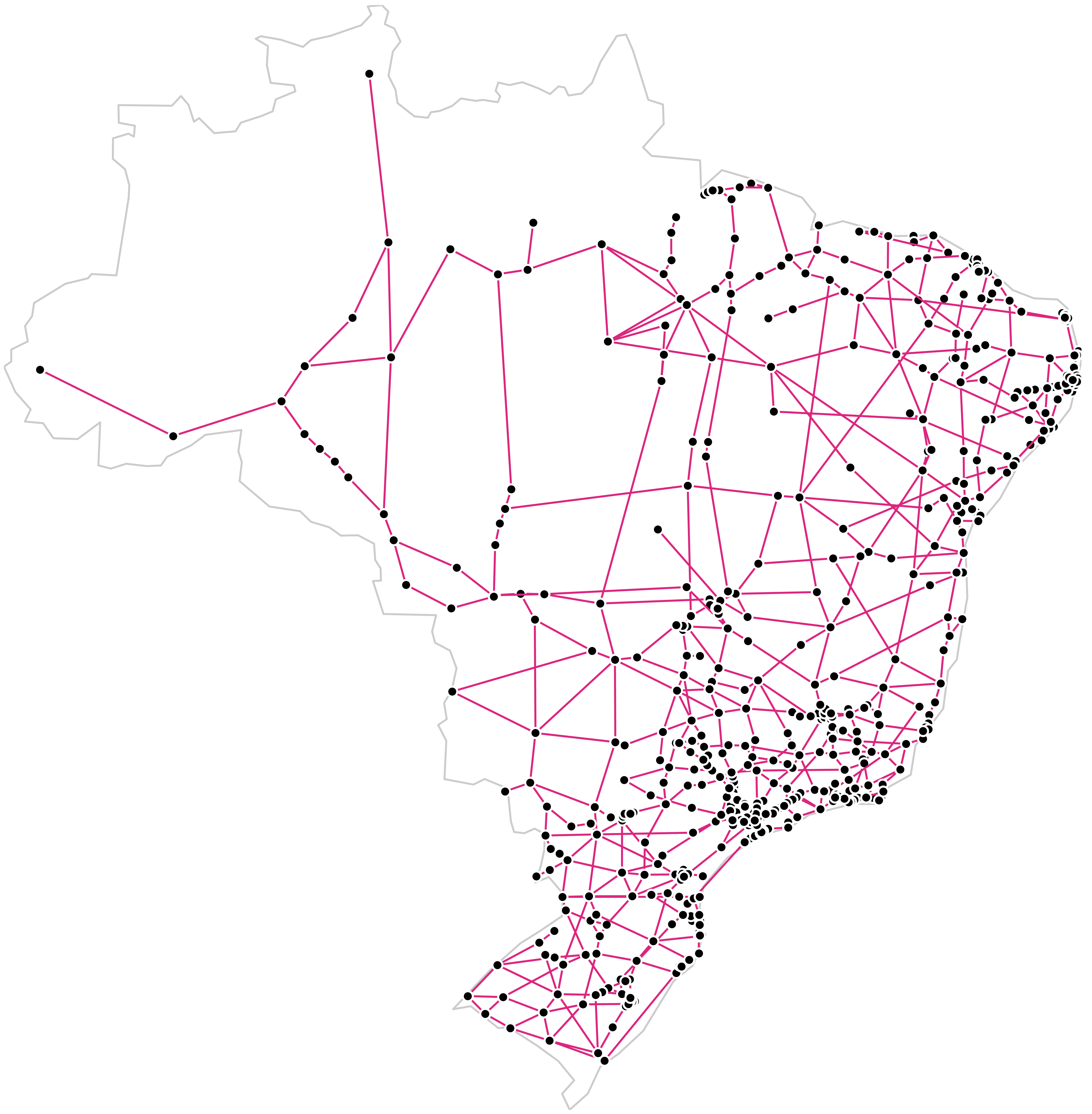}\\
    \includegraphics[width=0.48\linewidth]{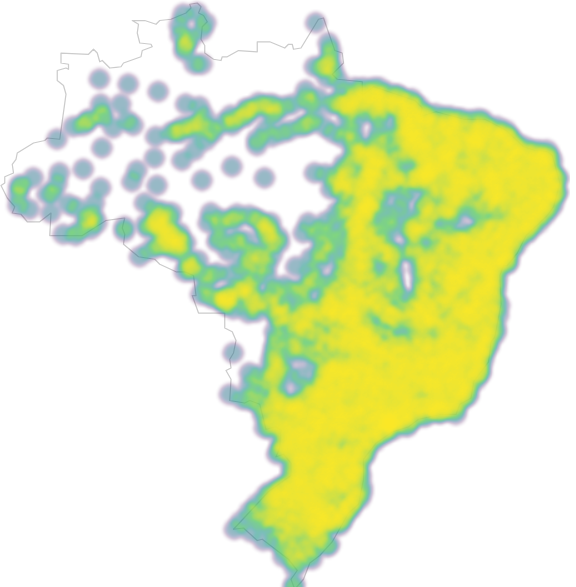}
    \caption{{\bf Brazilian Federal Road Network.} We see on the (top-left) the picture showing the road types in different colours as defined in Table \ref{tab:name}. However, some of these roads are not connected to others, so we use a modelling based on the 546 cities in Brazil and the federal roads that connected them leading to the giant component in the (top-right). Last, we show the population distribution of Brazil (bottom); note the concentration around the east coast.}
    \label{fig:map}
\end{figure}

The Brazilian federal road infrastructure comprises approximately 75,000 kilometres of roads \cite{PRF_Anuario_2023}. These are named according to a system that reflects their general direction and geographic location, as shown in Table \ref{tab:name}. The roads are depicted in Figure \ref{fig:map} (left), and the colours represent the different types of roads.

\begin{table}[ht]
\centering
\caption{{\bf Brazilian Federal Road Naming Conventions}. They are based on where they start in the country, their direction, and function \cite{dnit2020}. The colours in the table refer to the map in Figure \ref{fig:map} (left) where all the federal roads are depicted.}
\label{tab:name}
\setlength{\tabcolsep}{.5em} % Adjust column separation
\resizebox{.9\textwidth}{!}{%
\begin{tabular}{l c c l l}
\toprule
Road Type & General Format & Colour & Example & Description \\
\midrule
Radial & BR-0XX & \includegraphics[width=1cm]{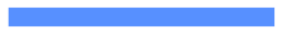} & BR-010, BR-020 & Roads from Bras\'ilia to country’s edges \\
Longitudinal & BR-1XX & \includegraphics[width=1cm]{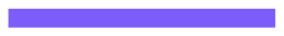} & BR-101, BR-116 & Roads oriented N-S\\
Transversal & BR-2XX & \includegraphics[width=1cm]{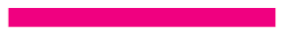} & BR-222, BR262 & Roads oriented E-W roads \\
Diagonal & BR-3XX & \includegraphics[width=1cm]{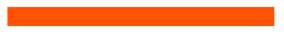} &BR-365, BR-319 & Roads oriented NW-SE or NE-SW \\
Connecting & BR-4XX & \includegraphics[width=1cm]{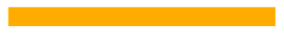}  & BR-407, BR488 & Generally connect federal roads\\
\bottomrule
\end{tabular}%
}
\end{table}

% \rmc{Julio, I need a description of the data here. What has been done describing period, what cleaning was done, what filtering, etc.}

For modelling the proposed networks, we used several datasets. The road structure was obtained from the National Road System (NRS) \cite{dnit_snv}, which provides georeferenced data for each kilometre of all Brazilian federal roads. From this dataset, we extracted information such latitude, longitude, and cities along the way. This dataset includes more than 130,000 geographic positions and was collected in 2022.

We also incorporated information about all 5,570 cities in Brazil, including their populations as collected in 2022 by the Brazilian Institute of Geography and Statistics (IBGE) \cite{ibge_censo_2022}. We utilised traffic incident data collected by the PRF, available through their open data portal \cite{prf_open_data}. This database contains over 485,000 incident records between 2017 and 2023, providing details such as date, time, number of people involved, injuries, location, to name a few.

For modelling the nodes, we selected cities classified by the IBGE as medium or large, with populations of at least 50,000 people \cite{ibge_censo_2022}. Brazil has a total of 5,570 cities, of which 3,126 are intersected by a federal road. The 544 selected cities have approximately 125 million residents, representing 76\% of the 164 million people residing in all cities traversed by federal road. In order to generate a consistent, we included an additional 30 small cities that are important for connecting roads (\eg, crossing points), resulting in a total of 574 cities. The nature of the Brazilian road network system, is that some of these cities are isolated and not part of the connected network of cities. Hence, we are left 546 cities which form the basis of the \rn (and variations); the network giant component. 

The edges in our network represent the roads linking the 546 selected cities. We utilised the NRS data, which contains information for each kilometre of road, to calculate the distances, number of cities, population, and incidents between pairs of cities. This calculation includes the end point as part of the weight; for instance, the population of people living between city A and city B, includes the population of A and B. 
% \rmc{Julio, eu não vou mudar isso agora, mas eu acho que se aceito, na versão final, precisamos pegar a população de uma cidade e dividir igualmente para todas as rodovias que a conectam} 
% For reference, we used the point on the road closest to the city centre to calculate distances. 
Figure \ref{fig:map} (centre) illustrates the road network, more specifically the giant component of the network which we call \rn.

We consider different weights in the structural analysis in this paper, giving rise to four weighted networks:

\begin{itemize}
    \item \rn: The road network using the distance as weight of the edges. This is essentially the network show in Figure \ref{fig:map} (centre).
    \item \crn: The same nodes as \rn but using the number of cities between the nodes as weight of the edges. For example, if between cities $A$ and $B$ we have $N$ cities, the weight of the edge $w(A,B) = N+2$ because the end points count for the weight.
    \item \prn: Weights here represent the total population in the cities in the \crn for that particular segment.
    \item \irn: Weights here represent the number of incidents that happened in-between the cities for the entire period of the study.
\end{itemize}

Table \ref{tab:network_measures} shows some basic measurements on the four networks. The number of nodes ($n$), edges ($m$), and average degree ($\langle k \rangle$) do not change, and hence it is only shown for the \rn. All the other measures are specific for each network.

\begin{table}[ht]
\begin{center}
\caption{{\bf Basic Measures of the Brazilian Federal Road Network.} We see the values for the exponent of the power law weighted distribution for each network ($\gamma_w$), the average shortest path ($\langle \ell_w \rangle$), diameter ($D_w$), and the weighted modularity ($Q_w$). Decimal places are not displayed for $\langle \ell_w \rangle$ and $D_w$ while for $n$, $m$, and $\langle k \rangle$ the values are shown once because they repeat for all networks.}
\label{tab:network_measures}
\setlength{\tabcolsep}{1em} % Adjust column separation
\resizebox{\textwidth}{!}{
    \begin{tabular}{lrrrr}
    \toprule
    Metric & \rn & \crn & \prn & \irn \\
    \midrule
    Number of Nodes ($ n $) & 546 & $\circ$ & $\circ$ & $\circ$ \\
    Number of Edges ($m$) & 761 & $\circ$ & $\circ$ & $\circ$ \\
    Average Degree ($\langle k \rangle$) & 2.78 & $\circ$ & $\circ$ & $\circ$\\
    \midrule
    Weighted Distribution Exponent ($\gamma_w$) & 1.78 & 1.76 & 1.73 & 1.45\\
    Average Shortest Path ($\langle \ell_w \rangle$) & 1,784 & 75 & 4,324,144 & 130 \\
    Diameter ($D_w$) & 5,768  & 213 & 22,902,144 & 45,659 \\
    Modularity ($Q_w$) &  0.852 & 0.842 & 0.867 & 0.887 \\
    \bottomrule
    \end{tabular}
    }
\end{center}
\end{table}

\section{Characterisation and Results}

The goal in this paper is to characterise the Brazilian road network, leading to a better clarity of the infrastructure. Yet, there are contextual information that need to be mentioned because of the four weight variations used. In the \rn the network can be used to understand hubs, effective paths, and more importantly how changes to this can lead to better efficiency in transportation times and fuel consumption; distance is highly correlated to travel time. In the \crn, one can concentrate on the identification of urban corridors as a function of the settlements, emergency planning, and planning of road maintenance making sure most cities remain connected at all times. A variation of this is the \prn where population of the cities are considered instead. In the \prn the analysis of paths (shortest) can demonstrate areas that are less populated and hence routes that may be maintained less often. Last, and relatively different from the others, is the \irn where the weights represent the number of incidents (\ie, deaths, injuries, driving under the influence). However, the specific applications described above fall outside the scope of this paper. Here we focus on identifying structural properties of these networks. 

\subsection{Weighted Degree Distributions}

In spatial networks, nodes are embedded in physical space, and connections between them are constrained by geographical proximity. This spatial embedding imposes limitations on the number of connections a node can have, leading to degree distributions that are generally more homogeneous and do not follow a power-law distribution typically seen in other types of complex networks \cite{barthelemy2011spatial,gastner2006spatial}. 
% The emergence of hubs with exceptionally high degrees is less likely because creating long-distance connections is costly and impractical in physical space.

When weights are assigned to the edges of spatial networks; representing attributes such as distances, the distribution of weighted degrees (also known as node strengths) can provide additional insights into the network’s structure and functionality. With weighted edges, the degree distributions in spatial networks like road networks may follow power laws \cite{barrat2004architecture} defined as $P(k) = C k^{-\gamma_w}$, where $k$ represents the weighted degree of a node. For our networks, the power law fitting leads to values of $1 \leq \gamma_w \leq 2$ which shows that the networks are extremely hub-dominated, which is generally a sign of vulnerability.

In the context of the \rn, examining the weighted degree distribution is essential for several reasons. When weights represent distances between cities, the distribution can highlight central nodes that are geographically significant due to their numerous or lengthy connections and perhaps some level of isolation given the other cities considered (with significant population) are far away. 
% In fact, the larger cities tend to have small values because they are surrounded by other also large cities do not feature on the max of \rn because they are surrounded by other large cities making the distances relatively small. 
If weights represent the number of intermediate cities (\crn), the distribution uncovers nodes that are endpoints of urban corridors, indicating potential areas of high socio-economic activity or congestion. When considering weights as the total population served by each segment (\prn), the weighted degree distribution helps identify cities that are crucial for connecting large populations, or cities for which a large fraction of the population depend on the road as an economic drive. Lastly, with weights representing the number of incidents (\irn), the distribution can pinpoint nodes that are hotspots for accidents, guiding targeted safety interventions and resource allocation to ensure the police can quickly respond to these incidents. 

Table \ref{tab:cities} shows the 5 cities with the maximum and minimum weights. It is worth noting the 2 largest cities in Brazil are on the highly ranked in the \prn, which happens because of the artifact that their population counts for the weight of the edges they have. In fact, we see two other cities are part of the great S\~ao Paulo: Osasco and Guarulhos. Other points to observe is the correlation between the \rn and \crn. We found that cities that have long distance roads attached to them are also the ones with several intermediate cities. Last, we see that the cities featuring in high ranks in the \irn are not the ones with high population weight, leading to the idea that incidents may be related to other issues such as road quality and geographical accidents.

\begin{table}[ht]
\centering
\caption{\textbf{Top 5 Cities with Maximum and Minimum Weights.} We can observe that, as expected, large metropolitan areas rank high in the \prn and they do not correlate directly to incidents in the \irn. However, there is some correlation to the list in \rn and \crn saying that cities that have long distance weights are similar to the ones that have many cities as intermediate nodes.}
\label{tab:cities}
\setlength{\tabcolsep}{0.5em} % Adjust column separation
\renewcommand{\arraystretch}{1.2} % Adjust row separation
\resizebox{\textwidth}{!}{ % Resize table to fit page width
\begin{tabular}{c*{8}{l}}
\toprule
\multirow{2}{*}{Rank} & \multicolumn{2}{c}{\rn} & \multicolumn{2}{c}{\crn} & \multicolumn{2}{c}{\prn} & \multicolumn{2}{c}{\irn} \\
\cmidrule(lr){2-3} \cmidrule(lr){4-5} \cmidrule(lr){6-7} \cmidrule(lr){8-9}
& Max & Min & Max & Min & Max & Min & Max & Min \\
\midrule
1 & Barreiras & Bel\'em 
 & Barreiras & Vit\'oria 
 & Osasco  & Camocim 
 & Registro  & Rio Claro  \\
 
2 & Balsas& Vitória 
 & Campo Mour\~ao & Jandira 
 & S\~ao Paulo  & Tr\^es Passos
 & Piraquara  & Brusque \\
 
3 & Jacobina& Jandira 
 & Jacobina & Aracaju
 & Rio de Janeiro & Caldas Novas
 & Betim & Valença\\

4 & Gurupi& Itapevi 
 & Balsas & Florian\'opolis 
 & Guarulhos & Camet\'a 
 & Bras\'ilia & Cabo Frio\\
 
5 & Barra do Gar\c{c}as & Aracaju
 & Picos & Brusque
 & Brasília & Graja\'u
 & Lages & Camet\'a \\
\bottomrule
\end{tabular}
}
\end{table}

\begin{figure}[ht]
    \centering
    \includegraphics[width=\textwidth]{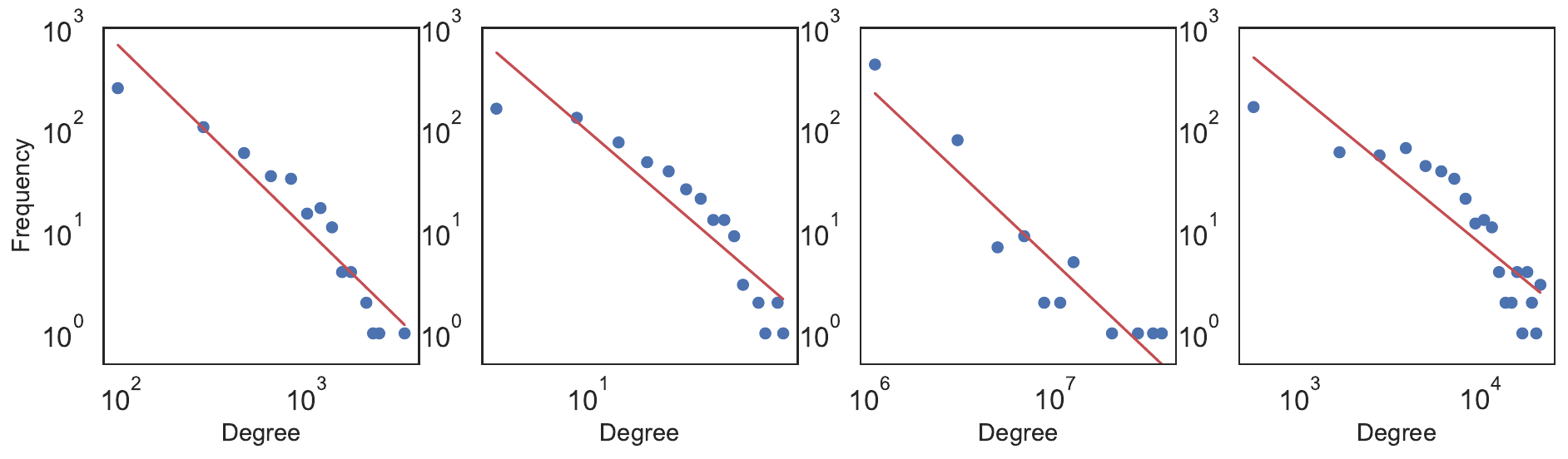}
    \caption{{\bf Weighted Degree Distributions.} The charts are for (from left to right): \rn, \crn, \prn, and \irn. It's worth noting that the distributions follow a power law demonstrating super-hubs. However, except for the \prn~\cite{soo2005zipf}, the networks appear to have some level of cut-off in their distribution which indicates limits to the values of the degrees.}
    \label{fig:distribution}
\end{figure}

Overall, understanding the weighted degree distribution allows for a holistic analysis of the network's topology and its implications. It helps in identifying critical nodes whose failure or inefficiency could disproportionately affect the network's functionality. This knowledge is vital for enhancing network robustness, optimizing traffic flow, improving safety measures, and supporting strategic planning and policymaking aimed at improving the transportation infrastructure. Figure \ref{fig:distribution} shows the degree distribution for each of the four networks considered.

While individual cases for cities are important in decision-making, understanding the distribution of these degrees is also fundamental to a holistic view of the vulnerability. Table \ref{sec:data} shows that the networks appear to be super-hub dominated. The distributions are depicted in Figure \ref{fig:distribution}

\subsection{Diameter and Paths}

The diameter of a weighted network, $D_w$ is defined as the longest shortest path between any pair of nodes, taking into account the weights assigned to the edges, as below:

\[
D_w = \max_{i, j \in n} d_{ij}^w,
\]

\noindent where \( d_{ij}^w \) is the weighted shortest path length between cities \( i \) and \( j \).

It represents the maximum cumulative weight that must be traversed to connect the most distant nodes in terms of the chosen weighting scheme. The diameter provides insights into the extent of the network's reach, potential bottlenecks, and areas where connectivity may be limited or require improvement. The values for all the networks are shown in Table \ref{tab:network_measures}.

In the context of the \rn, where the weights represent the distances between cities, the diameter signifies the greatest cumulative distance that must be covered to travel between the two most distant cities along the shortest possible route. This metric reflects the maximum physical separation in the network and can be used to assess the efficiency of national connectivity. Brazil, being a continental country with complex geographical features such as vast rivers, mountain ranges, and dense rainforests, inherently imposes lower bounds on these measures. For instance, the calculated diameter of the \rn is 5,768 kilometres, which exceeds the actual maximum straight-line distances across Brazil: 4,394 kilometres from the northernmost point at the Monte Caburaí in Roraima to the southernmost point at Arroio Chuí in Rio Grande do Sul. 

For the \crn, with weights representing the number of cities between each pair of connected cities, the diameter indicates the maximum number of cities that must be traversed along the shortest path between any two cities. This interpretation sheds light on the longest sequence of urban centres encountered on the most direct route, providing insights into urban density and regional development patterns. It can help identify corridors that connect numerous communities, potentially highlighting areas which could benefit from economic activity around transportation. For Brazil, there this diameter is 213 cities. This could be seen in conjunction to the actual population seen in the \prn, where weights denote the total population benefited by each segment, the diameter reflects the largest cumulative population connected along the shortest path between any two cities which for Brazil is about 22.9 million people which says that the diameter involves around 10\% of the entire Brazilian population (214 million people).

Last, in the \irn, the diameter represents the highest cumulative number of incidents encountered along the safest (least incident-prone) path between any two cities. This metric identifies the pair of cities for which even the safest route involves traversing segments with a high total number of incidents, highlighting potential areas of concern for road safety. The cumulative value for this diameter is more than 45 thousand incidents; indicating that some of the roads are quite dangerous, with about 18 incidents per day for the 7-year period studied. 

\begin{figure}[ht]
    \centering
    \includegraphics[width=0.4\linewidth]{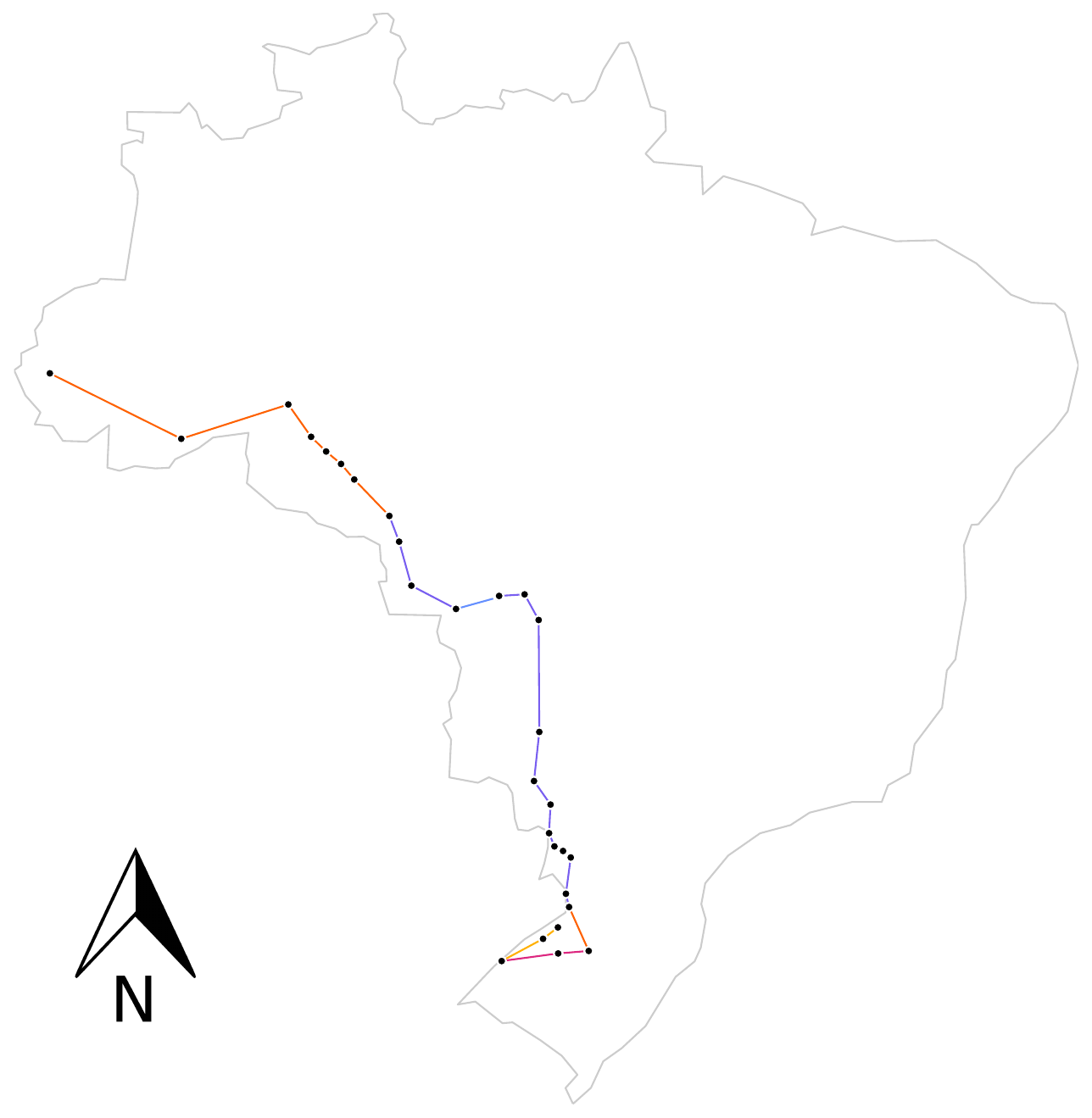}\hfill
     \includegraphics[width=0.4\linewidth]{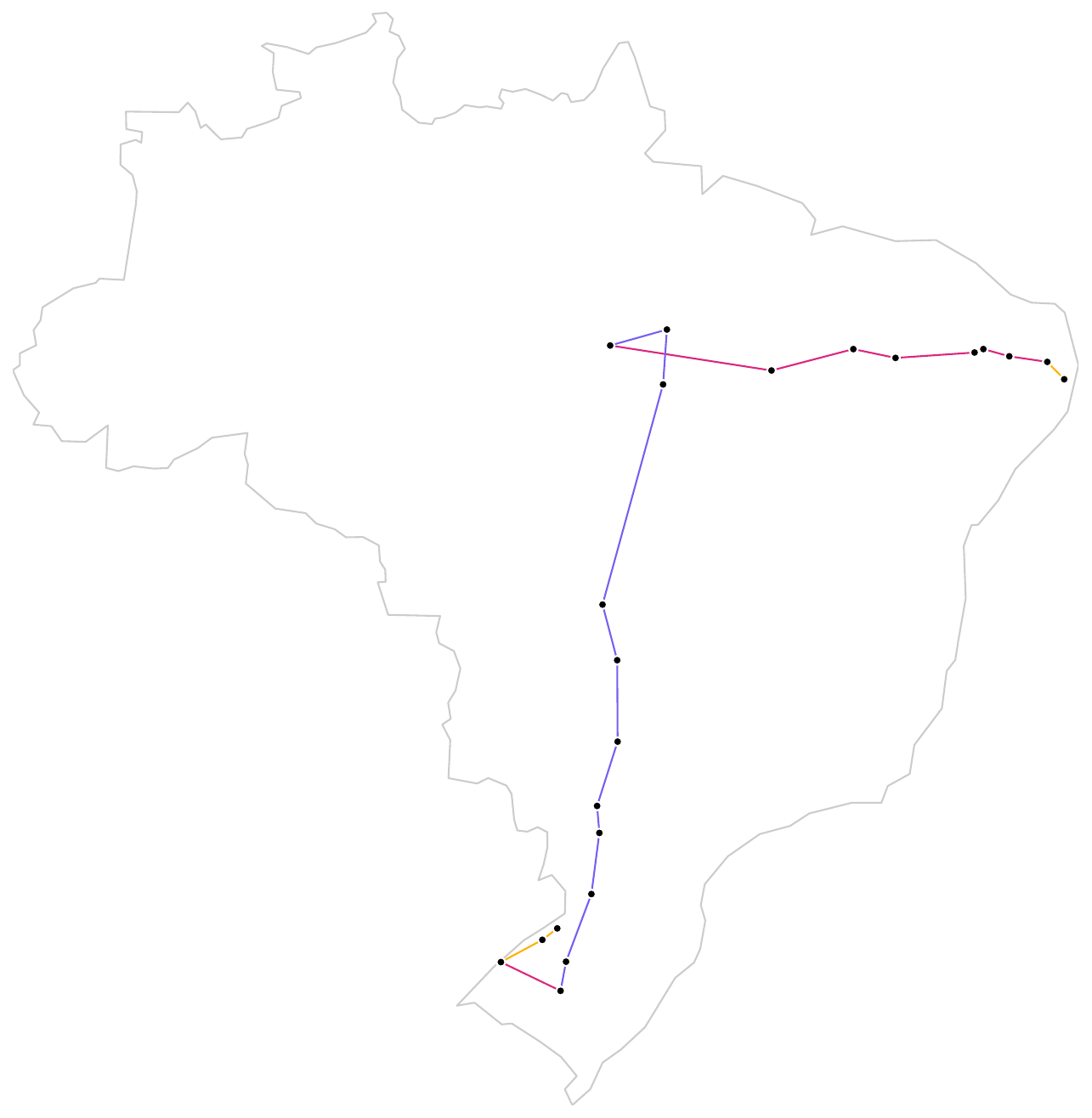}\\
    \includegraphics[width=0.4\linewidth]{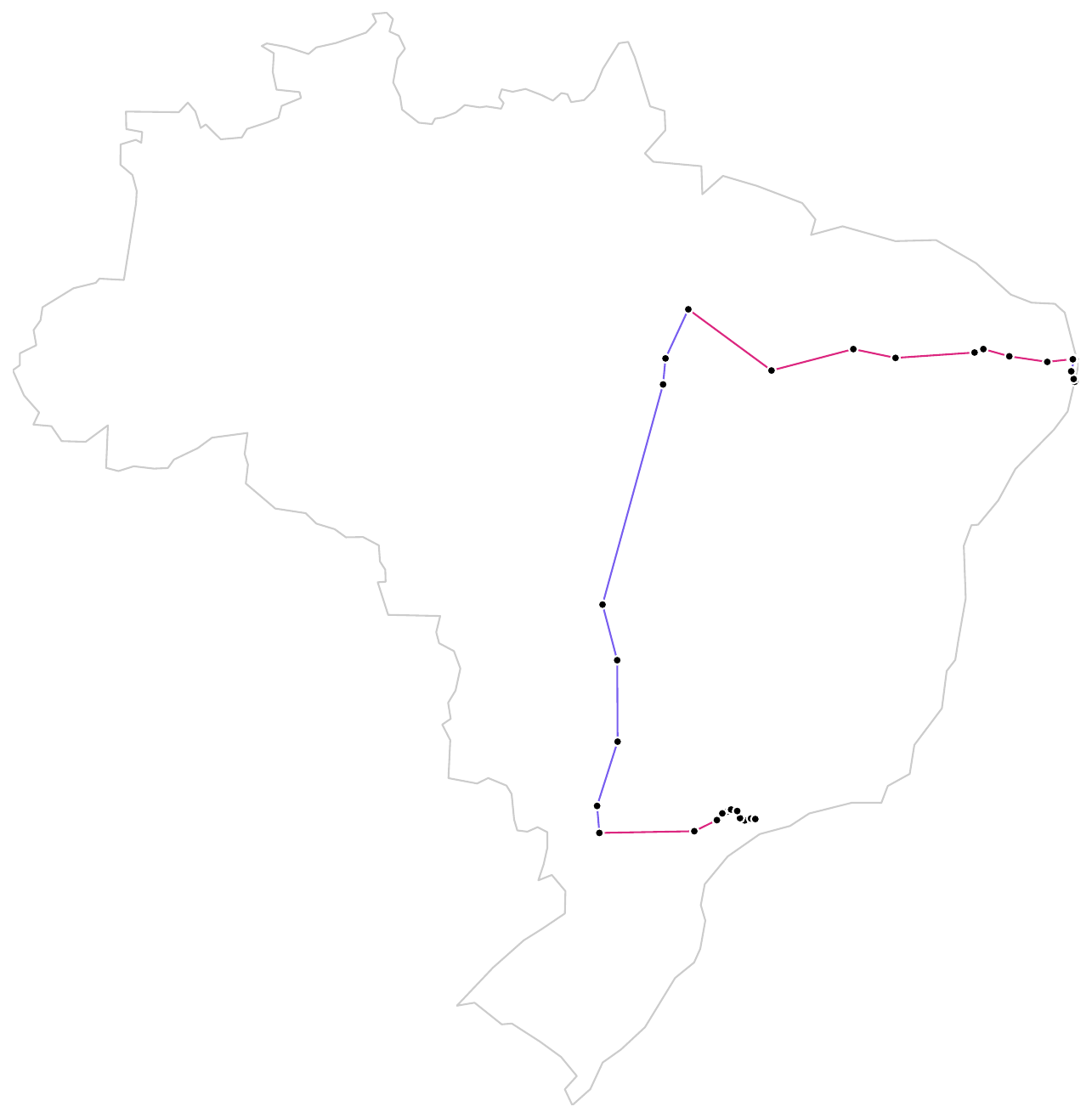}\hfill
    \includegraphics[width=0.4\linewidth]{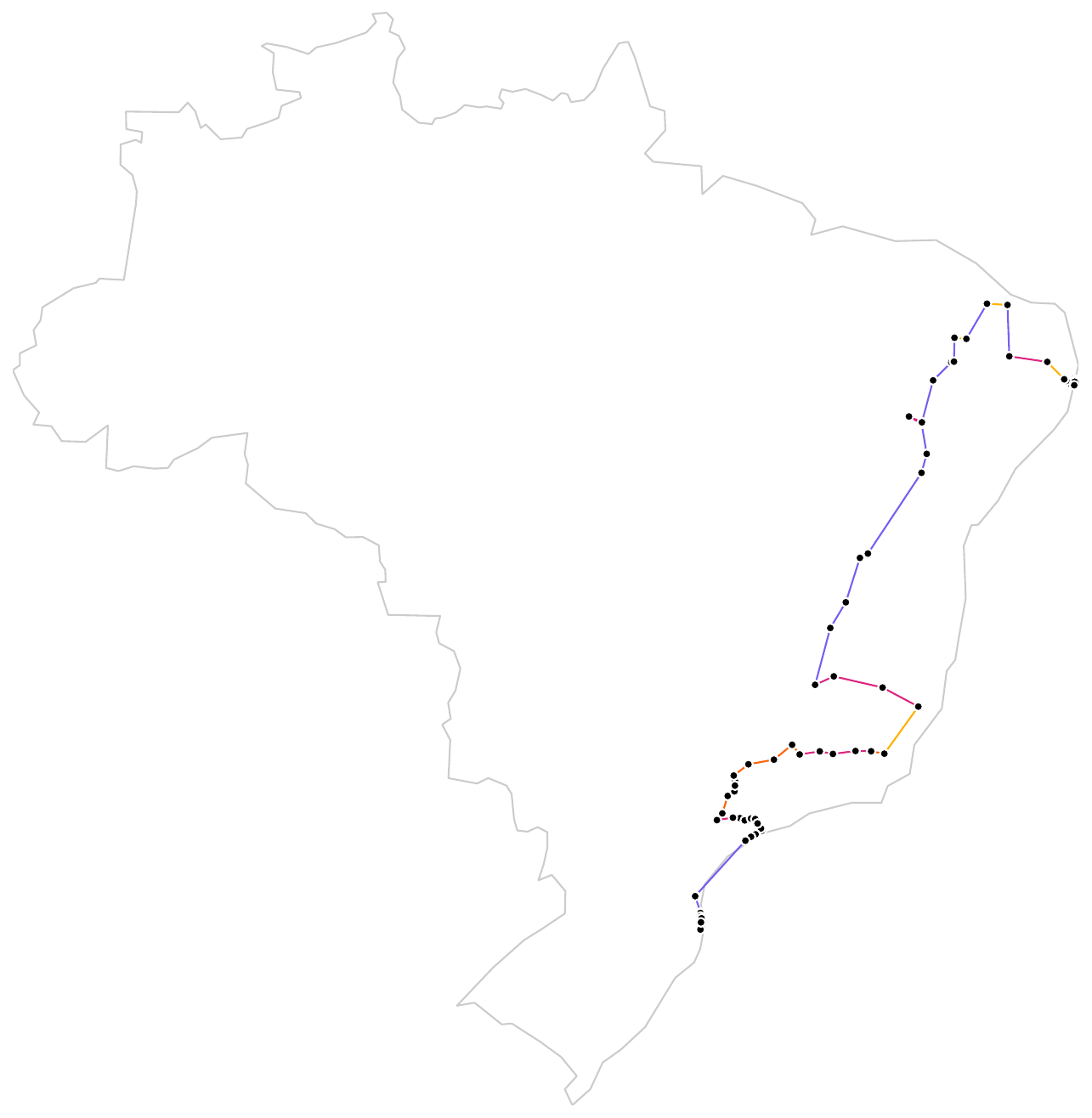}

    \caption{{\bf Diameters of the Networks.} The diameters of the \rn, \crn, \prn, and \irn (respectively left to right, top to bottom). The colours of the roads are used as defined in Table \ref{tab:name}.}
    \label{fig:diameter}
\end{figure}

Figure \ref{fig:diameter} shows the diameters for the four networks studied using the colours defined in Table \ref{tab:name}. Two things are worth noting, the significant difference between the \rn and the \irn, with the latter being concentrated on the more populated area of the country. The second is that there is a strong correlation between the \crn and the \prn with the main difference being that the \prn naturally gravitates to the city of S\~ao Paulo given its large population.

% \subsection{Centrality Measures}
% Centrality measures are used to identify the most important nodes in the network. We employ:
% \begin{itemize}
%     \item \textbf{Degree Centrality}: Identifies nodes with the highest number of direct connections.
%     \item \textbf{Betweenness Centrality}: Measures the importance of nodes that act as bridges between different parts of the network.
%     \item \textbf{Closeness Centrality}: Quantifies how easily a node can access all other nodes in the network.
% \end{itemize}

\subsection{Community Detection}

Community detection is a fundamental aspect of network analysis, aiming to uncover the underlying structure of a network by identifying groups (communities) of nodes that are more densely connected internally than with the rest of the network. This process is essential for understanding the modular organisation of complex systems, revealing how entities interact within and across different subgroups. 
% In various domains, such as sociology, biology, and infrastructure networks, community detection helps in identifying functional modules, social circles, or clusters of high interaction, providing insights that can inform targeted interventions, policy decisions, and optimisations.

Here we use the Louvain method \cite{blondel2008fast} which is a widely used algorithm for community detection due to its efficiency and ability to handle large and weighted networks. It operates by maximising the modularity of the network, in our case the weighted modularity, $Q_w$, a measure that quantifies the quality of a particular division of the network into communities \cite{newman2004analysis}. By considering edge weights, the Louvain method can detect communities that reflect not only the topology of the network but also the intensity of interactions between nodes (from weights). 

\begin{figure}[ht]
    \centering
    \includegraphics[width=0.4\linewidth]{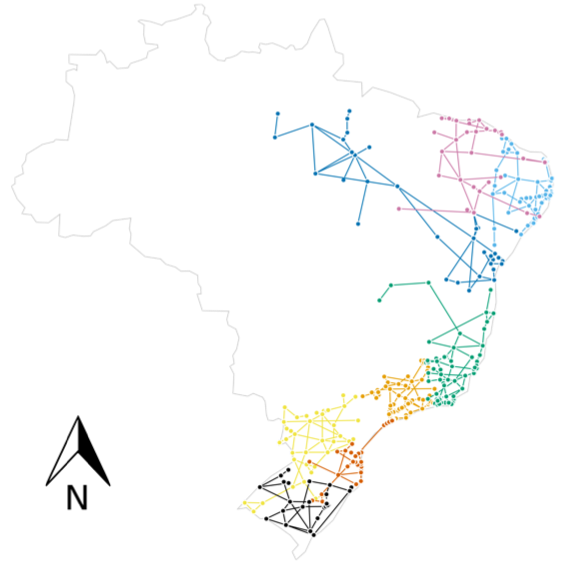}\hfill
     \includegraphics[width=0.4\linewidth]{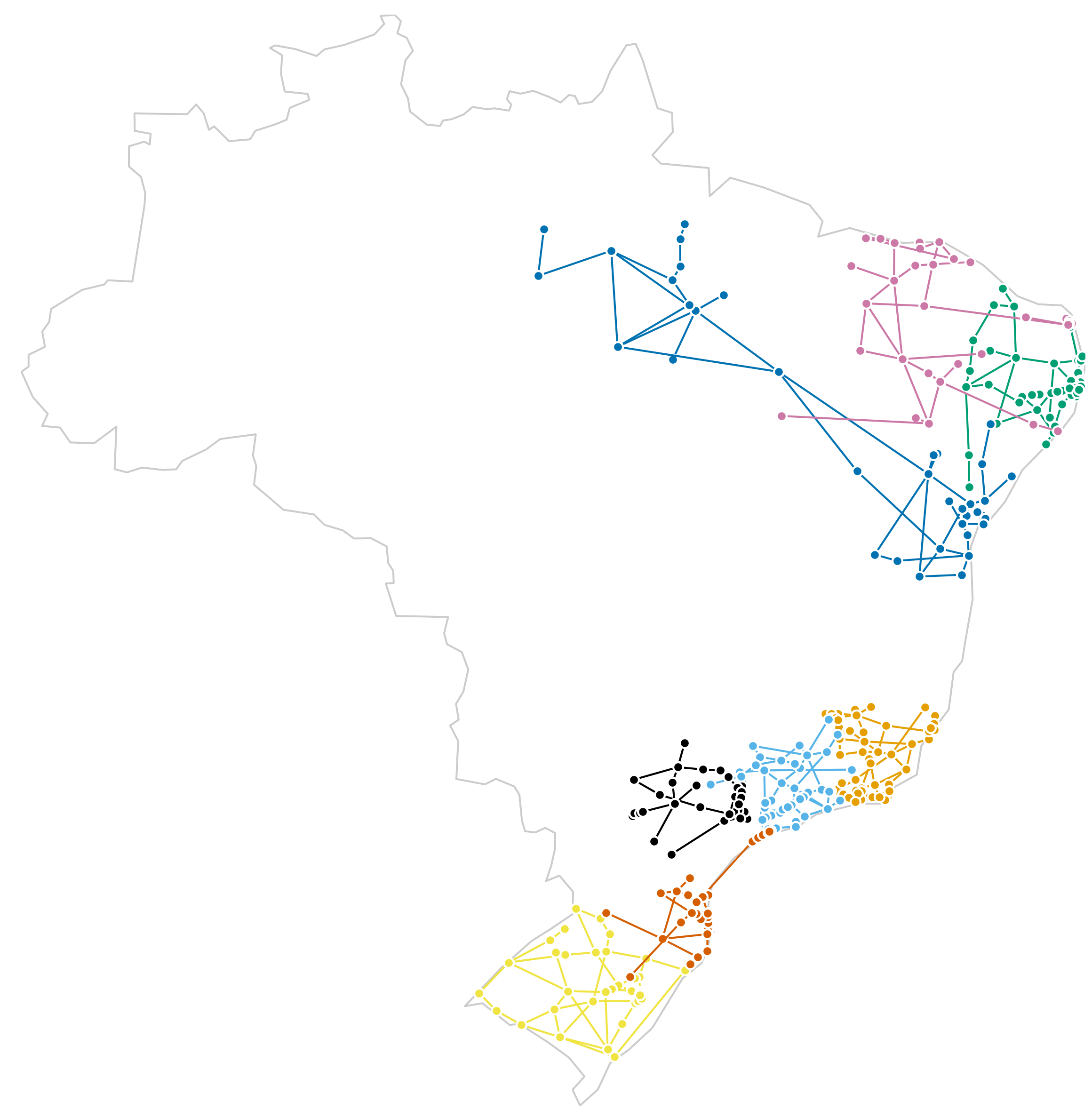}\\
    \includegraphics[width=0.4\linewidth]{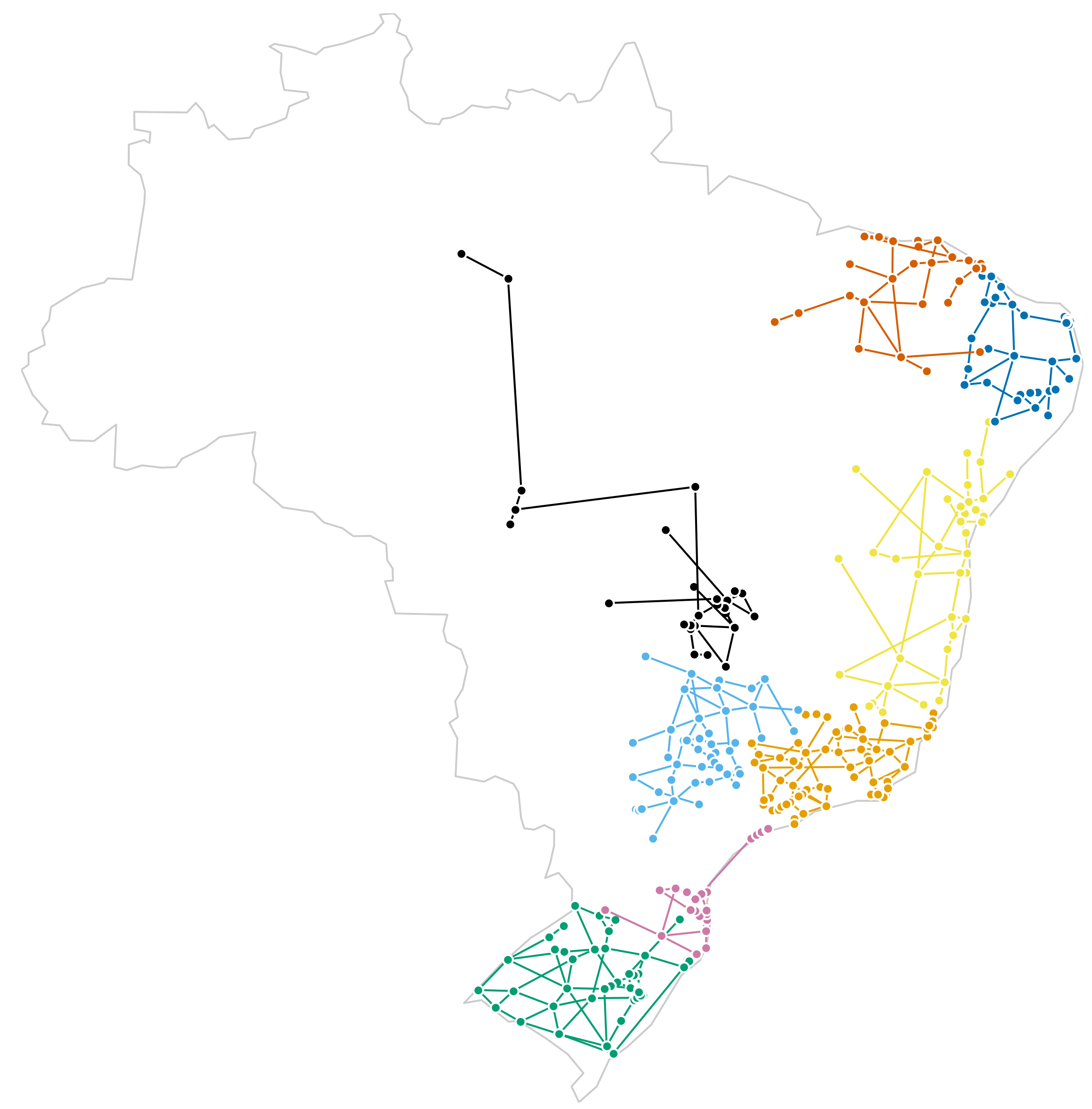}\hfill
    \includegraphics[width=0.4\linewidth]{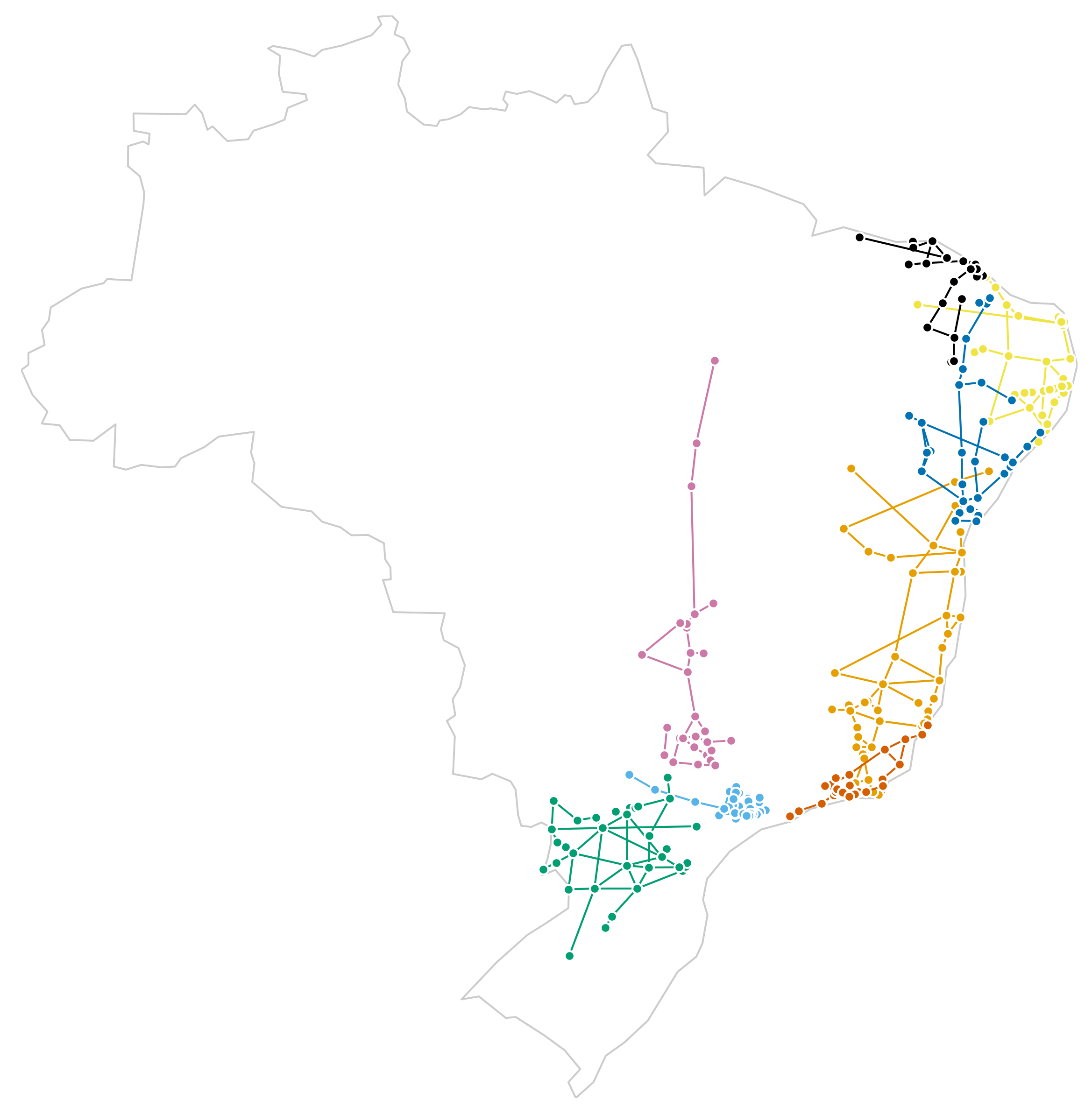}
    \caption{{\bf Network Communities.} For all networks, the communities show strong connections within regions. The picture shows the 8 largest communities for \rn, \crn, \prn, and \irn (from left to right, top to bottom).}
    \label{fig:community}
\end{figure}

In the case of the \rn and variations, the communities may correspond to regions where there is a high volume of traffic, densely connected urban areas, or zones with significant safety concerns due to a high number of incidents. 
% This weighted approach provides a deeper understanding of the network's modular organisation, which is crucial for effective planning, resource allocation, and policy-making.
%
Figure \ref{fig:community} shows the 8 largest communities for the four networks, respectively from left to right: \rn(19), \irn(26), \crn(20), \prn(20); the community analysis leads to more communities shown between parenthesis. One immediate aspect to observe is how the largest communities seem to be located in the coastal area of Brazil, where 55\% of the population lives (within 150km of the coast and 10\% of total territory) \cite{ibge_censo_2022}. 
% This concentration is due to historical settlement patterns, economic opportunities, and the presence of major urban centres such as São Paulo, Rio de Janeiro, Salvador, Recife, and Fortaleza \cite{ibge_censo_2022}, as presented in Figure \ref{fig:map}. 
% \rmc{I need some citations here. I found the information on the internet but not a document to cite like from IBGE.} \jtc{values updated with the 2022 census}

It is worth noting that the community structure on the networks is quite strong, as shown by the modularity values in Table \ref{sec:data}. Figure \ref{fig:community} depicts that the communities are quite regional, showing that the patterns of the nearby cities in the network have similarities. 

\section{On Network Resilience}

Resilience analysis, especially in the \rn, is critical due to its vital role in connecting cities across Brazil’s vast and diverse geography. The \rn as the backbone of Brazil’s transportation system, needs to show resilience to disconnections. Studying resilience allows for the identification of critical nodes (cities) whose removal, whether due to natural disasters, infrastructure failure, or other disruptions, could disproportionately impact the network. Natural and unnatural disasters, such as the devastating floods that recently impacted the southern region of Brazil, including the city of Porto Alegre, are likely to become more frequent and severe due to the growing effects of climate change. These events can lead to significant disruptions in critical infrastructure such as the \rn highlighting the need for proactive resilience planning. 
% By analysing the network’s resilience, we can identify the highly ranked nodes—those cities and connections that play a crucial role in maintaining overall connectivity.

Here, we compare node removals based on degree, weighted degree, and betweenness to random removal for it brings clarity to the structural importance of different nodes in the \rn. Degree, as a measure of the number of direct connections a node has, helps evaluate the impact of losing cities that are end points of several federal roads on the network’s overall connectivity. Weighted degree extends this analysis by incorporating the significance of these connections, using weights such as distance, population, or the number of intermediate cities. Betweenness centrality, on the other hand, assesses the role of cities as intermediaries or bridges, measuring how often a node lies on the shortest paths between other cities. This highlights nodes that are critical for maintaining the global flow within the network. Random removal serves as a null model to compare how the network responds when disruptions occur without targeting specific structural properties. 

\begin{figure}[ht]
    \centering
    \includegraphics[width=\linewidth]{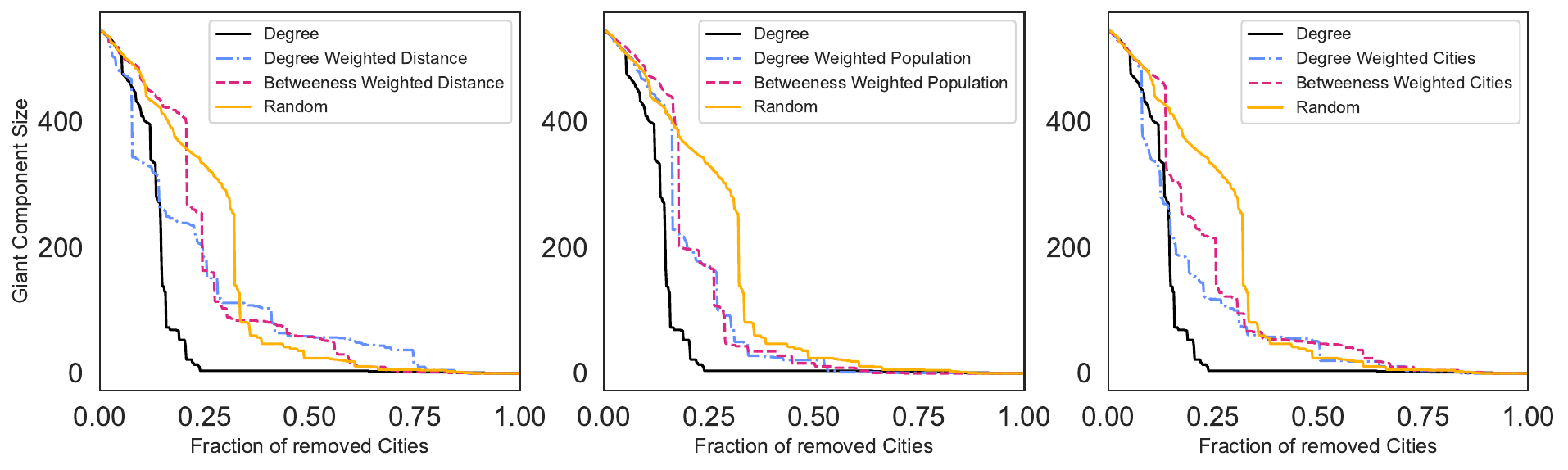}
    \caption{{\bf Resilience Analysis.} The impact of node (city) removals based on degree, weighted degree, and betweenness, compared to random, highlights the vulnerability of the giant component particularly for the removal of highly connected cities (degree). Here we see \rn, \prn and \crn respectively from left to right.}
    \label{fig:resilience}
\end{figure}

The results shown in Figure \ref{fig:resilience} (left) reveal that removing approximately 25\% of the nodes based on degree leads to the complete destruction of the giant component, with faster destruction observed in the \prn. This outcome suggests that cities with a high number of connections are crucial for maintaining large-scale connectivity, especially when these connections serve densely populated areas. The earlier impact of weighted degree removal in the \rn highlights that removing cities associated with large distances can disrupt connectivity also, but degree-based removals ultimately have a stronger and more widespread effect, possibly because we are dealing with spatial networks. Interestingly, the faster disintegration caused by degree removals in the \rn compared to betweenness suggests that, in spatial networks like the \rn, direct connections may be more critical than the intermediary roles captured by betweenness. This is due to the physical constraints and cost associated with forming long-distance connections, which make high-degree nodes less likely but disproportionately important.

\section{Discussion}
\label{sec:discussion}

This paper presented a comprehensive analysis of the \rn by leveraging network science to evaluate its structure, connectivity, and vulnerabilities. Four distinct network models were considered, with weights based on distance, population, number of intermediate cities, and incidents. These models provided a multifaceted perspective on the \rn, allowing for a holistic understanding of how different aspects of it contribute to its overall functionality. By analysing the structural properties of these weighted networks, we gained valuable insights into the key features that define the network's robustness and the critical nodes that play a pivotal role in maintaining connectivity.

The analysis highlights the importance of adopting a holistic approach to evaluate the \rn, particularly in the context of decision-making by the Federal Highway Police (PRF). Limited resources require strategic priorities, and a comprehensive view of the network enables the identification of areas where interventions can have the greatest impact. For example, the \prn can guide decisions on resource allocation to highly populated regions, while the \irn informs strategies monitoring the segments and maybe making sure the incidents are not due to poor road maintenance. Similarly, \crn sheds light on the role of smaller urban areas in maintaining regional cohesion, offering opportunities to strengthen economic incentives to critical corridors that may otherwise be overlooked. 

While resilience analysis demonstrated weaknesses associated with targeted node removals, the overarching value lies in understanding the broader structure of the \rn. In the future, we could work to build on the foundation of this paper by integrating multilayer networks that include additional transportation modes, such as waterways, railways, and airports. Furthermore, the use of dynamic data, such as traffic flow patterns and natural disaster simulations, would enhance the ability to anticipate and mitigate potential disruptions.

\bibliography{bib}

\begin{thebibliography}{10}

\bibitem{akbarzadeh2018look}
Meisam Akbarzadeh, Soroush Memarmontazerin, and Sheida Soleimani.
\newblock Where to look for power laws in urban road networks?
\newblock {\em Applied Network Science}, 3:1--11, 2018.

\bibitem{aschauer1989public}
David~Alan Aschauer.
\newblock Is public expenditure productive?
\newblock {\em Journal of monetary economics}, 23(2):177--200, 1989.

\bibitem{badhrudeen2022geometric}
Mohamed Badhrudeen, Sybil Derrible, Trivik Verma, Amirhassan Kermanshah, and
  Angelo Furno.
\newblock A geometric classification of world urban road networks.
\newblock {\em Urban Science}, 6(1):11, 2022.

\bibitem{barbosa2018human}
Hugo Barbosa, Marc Barthelemy, Gourab Ghoshal, Charlotte~R James, Maxime
  Lenormand, Thomas Louail, Ronaldo Menezes, Jos{\'e}~J Ramasco, Filippo
  Simini, and Marcello Tomasini.
\newblock Human mobility: Models and applications.
\newblock {\em Physics Reports}, 734:1--74, 2018.

\bibitem{barrat2004architecture}
Alain Barrat, Marc Barthelemy, Romualdo Pastor-Satorras, and Alessandro
  Vespignani.
\newblock The architecture of complex weighted networks.
\newblock {\em Proceedings of the national academy of sciences},
  101(11):3747--3752, 2004.

\bibitem{barthelemy2011spatial}
Marc Barth{\'e}lemy.
\newblock Spatial networks.
\newblock {\em Physics reports}, 499(1-3):1--101, 2011.

\bibitem{blondel2008fast}
Vincent~D Blondel, Jean-Loup Guillaume, Renaud Lambiotte, and Etienne Lefebvre.
\newblock Fast unfolding of communities in large networks.
\newblock {\em Journal of statistical mechanics: theory and experiment},
  2008(10):P10008, 2008.

\bibitem{bottasso2021roads}
Anna Bottasso, Maurizio Conti, Paulo~Costacurta de~Sa~Porto, Claudio Ferrari,
  and Alessio Tei.
\newblock Roads to growth: The brazilian way.
\newblock {\em Research in Transportation Economics}, 90:101086, 2021.

\bibitem{PRF_Anuario_2023}
{Brazilian Federal Highway Police (PRF)}.
\newblock Annual report 2023.
\newblock Technical report, Ministério da Justiça e Segurança Pública,
  Brasília, DF, 2023.
\newblock Accessed: 2024-10-29.

\bibitem{prf_open_data}
{Brazilian Federal Highway Police (PRF)}.
\newblock Open data on traffic accidents from brazilian federal highway police,
  2024.
\newblock Ministry of Justice and Public Safety.

\bibitem{ibge_censo_2022}
{Brazilian Institute of Geography and Statistics (IBGE)}.
\newblock Brazilian population census, 2022.

\bibitem{chalkiadakis2022urban}
Charis Chalkiadakis, Andreas Perdikouris, and Eleni~I Vlahogianni.
\newblock Urban road network resilience metrics and their relationship: Some
  experimental findings.
\newblock {\em Case Studies on Transport Policy}, 10(4):2377--2392, 2022.

\bibitem{chinazzi2020effect}
Matteo Chinazzi, Jessica~T Davis, Marco Ajelli, Corrado Gioannini, Maria
  Litvinova, Stefano Merler, Ana Pastore~y Piontti, Kunpeng Mu, Luca Rossi,
  Kaiyuan Sun, et~al.
\newblock The effect of travel restrictions on the spread of the 2019 novel
  coronavirus (covid-19) outbreak.
\newblock {\em Science}, 368(6489):395--400, 2020.

\bibitem{CNT2022}
{Confedera\c{c}\~ao Nacional do Transporte (CNT)}.
\newblock {Anu\'ario CNT do Transporte 2022 (in Portuguese)}, 2022.
\newblock Available at:
  \url{https://anuariodotransporte.cnt.org.br/2022/Inicial} (accessed: 25
  October 2024).

\bibitem{dnit2020}
{Departamento Nacional de Infraestrutura de Transportes}.
\newblock Nomenclatura das rodovias federais, 2020.
\newblock Accessed: 2024-10-29.

\bibitem{BrazilConstitution1988}
{Federal Republic of Brazil}.
\newblock {Constituição da República Federativa do Brasil de 1988}, 1988.
\newblock Available at:
  \url{https://www25.senado.leg.br/web/atividade/legislacao/constituicao-federal}
  (accessed: 25 October 2024).

\bibitem{ferrari2018economic}
Claudio Ferrari, Anna Bottasso, Maurizio Conti, and Alessio Tei.
\newblock {\em Economic role of transport infrastructure: Theory and models}.
\newblock Elsevier, 2018.

\bibitem{gastner2006spatial}
Michael~T Gastner and Mark~EJ Newman.
\newblock The spatial structure of networks.
\newblock {\em The European Physical Journal B-Condensed Matter and Complex
  Systems}, 49:247--252, 2006.

\bibitem{guimera2005worldwide}
Roger Guimera, Stefano Mossa, Adrian Turtschi, and LA~Nunes Amaral.
\newblock The worldwide air transportation network: Anomalous centrality,
  community structure, and cities' global roles.
\newblock {\em Proceedings of the National Academy of Sciences},
  102(22):7794--7799, 2005.

\bibitem{harvey2020condition}
David Harvey.
\newblock The condition of postmodernity.
\newblock In {\em The New social theory reader}, pages 235--242. Routledge,
  2020.

\bibitem{hulme2009trade}
Philip~E Hulme.
\newblock Trade, transport and trouble: managing invasive species pathways in
  an era of globalization.
\newblock {\em Journal of applied ecology}, 46(1):10--18, 2009.

\bibitem{kaluza2010complex}
Pablo Kaluza, Andrea K{\"o}lzsch, Michael~T Gastner, and Bernd Blasius.
\newblock The complex network of global cargo ship movements.
\newblock {\em Journal of the Royal Society Interface}, 7(48):1093--1103, 2010.

\bibitem{li2016world}
Zeyun Li and Sharifah Rohayah~Sheikh Dawood.
\newblock World city network in china: a network analysis of air transportation
  network.
\newblock {\em Modern Applied Science}, 10(10):213, 2016.

\bibitem{li2017road}
Zhigang Li, Mingqin Wu, and Bin~R Chen.
\newblock Is road infrastructure investment in china excessive? evidence from
  productivity of firms.
\newblock {\em Regional Science and Urban Economics}, 65:116--126, 2017.

\bibitem{liu2007small}
Cui-mei Liu and Jun-wei Li.
\newblock Small-world and the growing properties of the chinese railway
  network.
\newblock {\em Frontiers of Physics in China}, 2:364--367, 2007.

\bibitem{dnit_snv}
{National Department of Transport Infrastructure (DNIT)}.
\newblock National road system database (snv), 2024.
\newblock Ministry of Transport.

\bibitem{newman2003structure}
Mark~EJ Newman.
\newblock The structure and function of complex networks.
\newblock {\em SIAM review}, 45(2):167--256, 2003.

\bibitem{newman2004analysis}
Mark~EJ Newman.
\newblock Analysis of weighted networks.
\newblock {\em Physical Review E—Statistical, Nonlinear, and Soft Matter
  Physics}, 70(5):056131, 2004.

\bibitem{rodrigue2020geography}
Jean-Paul Rodrigue.
\newblock {\em The geography of transport systems}.
\newblock Routledge, 2020.

\bibitem{sen2003small}
Parongama Sen, Subinay Dasgupta, Arnab Chatterjee, PA~Sreeram, G~Mukherjee, and
  SS~Manna.
\newblock Small-world properties of the indian railway network.
\newblock {\em Physical Review E}, 67(3):036106, 2003.

\bibitem{skidmore2009brazil}
Thomas~E Skidmore.
\newblock Brazil: Five centuries of change.
\newblock {\em OUP Catalogue}, 2009.

\bibitem{soo2005zipf}
Kwok~Tong Soo.
\newblock Zipf's law for cities: a cross-country investigation.
\newblock {\em Regional science and urban Economics}, 35(3):239--263, 2005.

\bibitem{tak2018measuring}
Sehyun Tak, Sunghoon Kim, Young-Ji Byon, Donghoun Lee, and Hwasoo Yeo.
\newblock Measuring health of highway network configuration against dynamic
  origin-destination demand network using weighted complex network analysis.
\newblock {\em PLoS one}, 13(11):e0206538, 2018.

\bibitem{weber2012evolving}
Joe Weber.
\newblock The evolving interstate highway system and the changing geography of
  the united states.
\newblock {\em Journal of Transport Geography}, 25:70--86, 2012.

\end{thebibliography}
\bibliographystyle{plain}

\end{document}